\documentclass[preprint,1p,11pt]{elsarticle}
\usepackage{mathrsfs}
\usepackage{epsfig}
\usepackage{amsmath}
\usepackage{amssymb}
\usepackage{slashed}

\newcommand{\bea}{\begin{eqnarray}}
\newcommand{\eea}{\end{eqnarray}}
\newcommand{\bnn}{\begin{eqnarray*}}
\newcommand{\enn}{\end{eqnarray*}}
\newcommand{\be}{\begin{equation}}
\newcommand{\ee}{\end{equation}}

\journal{academic journal}
\begin{document}

\begin{frontmatter}
\title{Conformal Relativity with Hypercomplex Variables}

\author{S. Ulrych}
\address{Wehrenbachhalde 35, CH-8053 Z\"urich, Switzerland}

\begin{abstract}
Majorana's arbitrary spin theory is considered in a hyperbolic complex representation.
The underlying differential equation is embedded into the gauge field theories of Sachs and Carmeli. In particular, the 
approach of Sachs can serve as a unified theory of general relativity and electroweak interactions.
The method is extended to conformal space with the intention to introduce the strong interaction.
It is then possible to use the wave equation, operating on representation functions of the conformal group, to describe the dynamics of matter fields.
The resulting gauge groups resemble closely the gauge symmetries of Glashow-Salam-Weinberg
and the Standard Model.
\end{abstract}

\end{frontmatter}

\section{Introduction}
\label{intro}
Relativistic quantum physics can
be founded on the Dirac equation or a Klein-Gordon equation with spin, as both equations can be formulated in a way that they include
the same mathematical content. On the one hand one has to deal with the simple structure of a first
order differential equation. The price to pay is matrix structures which are more complex than necessary. 
On the other hand there is the higher
complexity of a second order differential equation, which is accompanied by a simplification in the applied matrix algebra.
Feynman had a preference for the Klein-Gordon fermion equation as the path integral formalism can be 
handled more easily in this representation \cite{Fey58}.
The Klein-Gordon fermion equation, which is equivalent to the squared Dirac equation, has been initially derived by Kramers \cite{Kra33,Kra57} and Lanczos \cite{Lan33}. One could denote Klein-Gordon equations with spin, which are congruent to this representation, as Dirac-Klein-Gordon equations.
The hyperbolic complex second order differential equation introduced in \cite{Ulr05} belongs to this category.

The Dirac equation and its quadratic counterpart are used for spin $1/2$ particles.
Furthermore, it is possible to formulate wave equations of arbitrary spin.
This has been done by Majorana \cite{Maj32}, Dirac \cite{Dir36}, Fierz and Pauli \cite{Fie39a, Fie39}, 
Gelfand and Yaglom \cite{Gel48}, Nambu \cite{Nam66,Nam67}, Fronsdal \cite{Fro67}, Gitman and Shelepin \cite{Git01}, and others. 
Consider in this context also the review articles of Fradkin \cite{Fra66} and Esposito \cite{Esp12}.
An important feature of the work of Majorana, Gelfand and Yaglom is the resulting mass spectrum in dependence of the spin
of the considered quantum state.

Recently, this mass spectrum could be reproduced with an ordinary d'Alembert operator 
acting on representation functions of the Poincar\'e group \cite{Ulr13}.
The result is a generalized Klein-Gordon equation, where a 
parameter that represents spin appears beside the mass parameter as part of the fundamental single particle field equation.
One could denote this equation as Majorana-Klein-Gordon equation in order to distinguish from the Dirac-Klein-Gordon equation mentioned above.
In the Majorana-Klein-Gordon equation aspects of the underlying relativistic 
Clifford algebra are moved to the solution space, whereas the differential operator itself
is free of spin representations.

The spectrum of the d'Alembert operator, when applied to the representation functions of the Poincar\'e group, gives rise to three parameters.
One of them can be considered as a constant coefficient,
which induces a relation between mass and spin in analogy to investigations of Gelfand and Yaglom \cite{Gel48}, 
Barut et al. \cite{Bar68b} and Varlamov \cite{Var12}.
In fact one has the choice, which parameters are considered to be physical as the 
interpretation of the constant is still unclear.
Based on this choice one can reinterpret
momentum and mass of the Poincar\'e group representation functions as given in \cite{Ulr13} and see them only as coefficients
in a series expansion. What is considered as energy, momentum, and mass has to be derived at a later stage,
for example in correspondence with the Fourier transformation of the Green function of the considered system.
More details can be found in Section~\ref{wave}.

Sachs developed a unified theory of gravitation and electroweak interactions with a Maxwell-like representation of 
Einstein's general relativity \cite{Sac82, Sac10}. 
The method is based on a quaternionic paravector algebra, 
which is congruent to the hyperbolic complex Pauli algebra applied in \cite{Ulr13}.
Sachs states that the Majorana equation is the most general irreducible expression of relativistic quantum mechanics, in special relativity.
It is thus
natural to introduce the Majorana-Klein-Gordon equation in the representation of \cite{Ulr13} into the Sachs framework
of quaternionic physics. The notion of a quaternion has to be understood now in a generalized sense
as it covers hyperbolic complex arbitrary spin representations. 
Consider here Sections~\ref{matter}, \ref{wave}, \ref{general}, and \ref{interaction}.

The conformal group as a generalization of the Poincar\'e group
has been applied extensively to describe physical systems, see the review article of Kastrup \cite{Kas08}.
In the sixties of the last century there was the hope that the new experimental observations in nuclear 
and particle physics could be explained with the help of conformal symmetries. One may consider for example
publications of Kastrup \cite{Kas62}, Hepner \cite{Hep62}, Mack and Salam \cite{Mac69}, Wess \cite{Wes60a, Wes60b}, or Flato et al. \cite{Fla70}.
Attempts to describe the baryon spectrum in terms of the conformal group have been published by
Nambu \cite{Nam66,Nam67} and Barut and Kleinert \cite{Bar67}. 
Meson decay rates and proton-proton scattering observables
have been calculated in good agreement with experiment in a conformal model
by Barut, Tripathy, and Corrigan \cite{Bar67b,Bar68}. Furthermore, the conformal group does not seem to be of relevance only in the context of the
strong interaction, but also with respect to cosmology.
It has been found that the conformal invariance produces a good correlation to experimental data of
the cosmological redshift, see Segal \cite{Seg76,Seg93}. 
In addition, a conformal extension of Einstein's general relativity has been applied by Mannheim 
in order to explain velocity curves of rotating galaxies \cite{Man06}.

Thus there are arguments to apply the conformal compactification to Minkowski space 
to obtain a conformal hyperbolic complex paravector algebra.
It is then possible to extend the theory of Sachs to conformal space with the 
intention to incorporate the strong interaction. 
The hyperbolic complex conformal algebra is discussed starting from Section~\ref{confalg}.
From the perspective of general relativity the extension is
related to the twistor approach of Penrose \cite{Pen67} and the supergravity theories of
Freedman et al. \cite{Fre76, Fre12} and Volkov and Soroka \cite{Vol73,Vol74}.
Consider in this context also the twistor string theory of Witten \cite{Wit04}.
The relation to these approaches points also into the direction of the anti-de Sitter/conformal field theory (AdS/CFT) correspondence
of Maldacena \cite{Mal98}, Gubser et al. \cite{Gub98}, and Witten \cite{Wit98}.

\section{The Hyperbolic Pauli Algebra in Minkowski Space}
\label{para}
The paravector model of Minkowski space based on complex quaternions
has been applied to relativistic physics for example by Sachs \cite{Sac82}, Baylis \cite{Bay99}, and Carmeli \cite{Car01}.
See Gsponer and Hurni for an extensive list of references in this research area \cite{Gsp08}.
The algebra of complexified quaternions corresponds to the real Clifford algebra $\mathbb{R}_{3,0}$. 
The product of two basis vectors of the form $e_\mu=(1,e_i)$, which include the three basis elements $e_i$ of $\mathbb{R}_{3,0}$,
can be used to define the equation
\be
\label{basis4}
e_\mu\bar{e}_\nu=g_{\mu\nu}-\imath \sigma_{\mu\nu}\,.
\ee
Here the conjugation anti-involution has been applied to the second element on the left hand side of the equation
in order to obtain the correct metric properties of Minkowski space.
Carmeli \cite{Car01} works without conjugation, but uses instead primed and unprimed indices 
for the matrices of the algebra,
which refer to the two distinct representations of a relativistic spinor.

The product of two basis elements has been separated in Eq.~(\ref{basis4}) into
symmetric and anti-symmetric contributions. The symmetric part defines the metric tensor
\be
\label{dotprod}
g_{\mu\nu}=e_\mu\cdot e_\nu=\frac{e_\mu\bar{e}_\nu+e_\nu\bar{e}_\mu}{2}\,.
\ee
The non-zero matrix elements of the metric tensor are
proportional to the identity element of the algebra. The identity is
represented in general in terms of a $n\times n$ matrix, where $n$ corresponds to the dimension 
of the spin representation~\cite{Ulr13}. 

The antisymmetric contributions in Eq.~(\ref{basis4}) define the spin tensor, which is given by the following expression
\be
\label{spinprod}
-\imath \sigma_{\mu\nu}=e_\mu\wedge e_\nu =\frac{e_\mu\bar{e}_\nu-e_\nu\bar{e}_\mu}{2}\,.
\ee
The matrix elements of the antisymmetric tensor are represented as $n\times n$ matrices. 
The tensor can be used to define the relativistic spin angular momentum operator
\be
\label{spin}
s_{\mu\nu}=\frac{\sigma_{\mu\nu}}{2}\,.
\ee
The explicit form of the antisymmetric tensor in terms of the basis elements and the pseudoscalar 
is displayed in \cite{Ulr13} with a different sign convention.

The pseudoscalar is interpreted in geometric algebra as a directed volume element or volume form,
see Doran and Lasenby \cite{Dor03}.
The pseudoscalar of the Clifford algebra $\mathbb{R}_{3,0}$ is introduced here
as the product of all basis elements of the paravector algebra
\be
\label{pseudo}
\imath=e_0\bar{e}_1e_2\bar{e}_3=ij\,.
\ee
In the hyperbolic complex representation of the algebra the pseudoscalar
includes the complex unit $i\equiv\sqrt{-1}$ and the
hyperbolic unit $j\equiv\sqrt{+1}$. 
This implies that quantum phases in a relativistic Hilbert space will have a more detailed structure 
in this geometric model than usual. 
It has been suggested also by other authors to apply different phase units  to 
modifications of standard quantum physics.
Pure hyperbolic quantum phases and their interferences have been investigated 
previously by Khrennikov, see \cite{Khr10,Khr10b} and the references therein. Phases of parabolic numbers 
with the unit $\epsilon\equiv\sqrt{0}$ have
been applied by Kisil to study the relation between classical and quantum mechanics~\cite{Kis12a}. 

The hyperbolic unit in the pseudoscalar can be traced back to the hyperbolic complex representation of
$\mathbb{R}_{3,0}$, where the basis elements are given by
\be
\label{Pauli}
e_i=j\sigma_i\,.
\ee
Here the Pauli algebra is multiplied by the hyperbolic unit.
The hyperbolic Pauli matrices can be considered as the two-dimensional fundamental representation of the real Clifford algebra $\mathbb{R}_{3,0}$.
The hyperbolic unit can be introduced into the Pauli algebra also by complexification, which has been demonstrated by Baylis and Keselica 
in order to analyze the structure of the Dirac equation \cite{Bay12}. 
Furthermore, Mironov and Mironov applied the tensor product of a Pauli paravector model
to reformulate differential equations in relativistic quantum physics \cite{Mir09,Mir13}.
Equation~(\ref{Pauli}) is not a complexification and not a tensor product as the algebra still generates only 
eight distinct elements and thus remains isomorphic to the Pauli algebra.
Thus the notion of a hyperbolic Pauli algebra can be introduced as a synonym for a hyperbolic complex representation of the Pauli algebra.
Complexified versions of the Pauli algebra are identified as the
sixteen element complex Clifford algebra $\bar{\mathbb{C}}_{3,0}$. The tensor product of the Pauli algebra
corresponds to the Dirac algebra $\mathbb{R}_{4,1}$.

Equation~(\ref{basis4}) can be considered from another perspective if one takes into account that a Clifford algebra
encodes the group manifold of the orthogonal transformations in the considered space. 
The basis elements of the Clifford algebra form a noncoordinate basis and Eq.~(\ref{basis4}) may be reformulated as
\be
\label{basis5}
e_\mu\bar{e}_\nu=g_{\mu\nu}-\imath c_{\mu\nu}^{\phantom{\mu\nu}\sigma} e_\sigma\,.
\ee
The second term on the right hand side of Eq.~(\ref{basis5}) encodes the structure constants of
the Lorentz group manifold. The explicit values can be read off by comparison with $\sigma_{\mu\nu}$.

\section{Matter Fields}
\label{matter}
Matter fields can be expanded in terms of plane wave representation functions of the
Poincar\'e group. In the terminology of \cite{Ulr13}  these representations were called quaternion waves.
They are labeled by 
\be
\label{irred}
\psi(x)\equiv \psi_{n\upsilon\tau}(x)=\langle x \vert n,\upsilon,\tau\rangle\,.
\ee
Here the integer value $n$ refers to the number of polarization states and 
$\upsilon$ is a four vector which has been identified with the momentum in \cite{Ulr13}, but 
following the discussion in Section~\ref{intro} the vector $\upsilon$ is interpreted now as an expansion coefficient.
The invariant obtained by the squared vector coefficient $\upsilon^2$ has been omitted in Eq.~(\ref{irred}).
The representation functions are in fact matrix valued. One may see them as a whole,
introduce additional indices referring  to the matrix elements, or append a spinor for comparison with standard spinor physics.
In the latter case one works with the vector bundle associated to the Poincar\' e group principle bundle.
The corresponding group transformations are then operating in spaces $\mathbb{R}^n$ of arbitrary spin.

The notation $\tau$ is referring  to quantum numbers of the CPT-transformations.
The transformations form the cyclic group 
$\mathbb{Z}_2\otimes\mathbb{Z}_2\otimes\mathbb{Z}_2$.
The representation which stands for $\tau\equiv\{+++\}$ may be chosen as
\be
\label{irred2}
\psi(x)=\exp{(-\imath x \bar{\upsilon})}\,.
\ee
Here coordinate vector, vector coefficients, and basis elements of the Clifford algebra
have been abbreviated by
\be
x\bar{\upsilon}=e_\mu\bar{e}_\nu x^\mu \upsilon^\nu\,.
\ee
Representations referring  to other values of $\tau$ can be obtained by changing the sign of the
exponent, the position of $x$ and $\upsilon$, and the position of the bar symbol.

For the following investigations it is useful to introduce an alternative
representation of the wave function, which includes a spin vector coefficient
\be
\label{planespinwave}
\psi(x)=\exp{(-\imath x\cdot(\upsilon+\varsigma))}\,.
\ee
The spin vector has been identified in \cite{Ulr13} with  the Pauli-Lubanski vector.
This interpretation changes in analogy to the momentum vector as the spin vector is derived from the expansion coefficient
$\upsilon$ of the representation functions
\be
\label{lubadef}
\varsigma_\mu=e_\mu\wedge \upsilon=
-\imath\sigma_{\mu\nu}\upsilon^\nu=\epsilon_{\mu\rho\sigma\nu}s^{\rho\sigma}\upsilon^\nu\;.
\ee
The coordinates of the spin vector are biparavectors, which can be interpreted geometrically as
planes formed by the basis vectors and the vector coefficient $\upsilon$. For more details on biparavectors
see Baylis \cite{Bay99}. Equation ~(\ref{lubadef}) displays furthermore how the spin vector coefficients are
related to the relativistic spin angular momentum operator.

The momentum operator is introduced as the derivative with respect to the space-time coordinates multiplied by the pseudoscalar of the algebra
\be
\label{momentum}
p_\mu= \imath \frac{\partial}{\partial x^\mu}\,.
\ee
If the momentum operator is applied to the wave function given by Eq.~(\ref{planespinwave}) one finds
\be
\label{eigen}
p_\mu\psi(x)=(\upsilon+\varsigma)_\mu\psi(x)\,.
\ee
This equation indicates, why vector and spin vector coefficients in the
Poincar\'e group representation functions are not identified with momentum 
and Pauli-Lubanski vector. 

The momentum operator as defined above can be used to perform a translation of the
plane wave
\be
\label{translation}
T(a)\psi(x)=e^{-\imath a\cdot p}\psi(x)=e^{-\imath a\cdot (\upsilon+\varsigma)}\psi(x)=\psi(x+a)\,.
\ee
The translation operator is thus acting on the Hilbert space function with the same overall result as 
in standard quantum mechanics.

Sachs developed a theory of elementary matter based on a complex quaternion algebra, which is congruent to the
algebra introduced in Section~\ref{para}.
With the principle of least action Sachs derives Maxwell-like field equations 
of general relativity from a quaternionic Einstein-Hilbert Lagrangian \cite{Sac67}. 
The basis elements of the quaternion frame of Sachs are affected by coordinate transformations
in the following way
\be
\label{trafo}
 \vartheta e_\mu \vartheta^\dagger=\frac{\partial \tilde{x}^\nu }{\partial x^\mu} e_\nu\,.
\ee
The dagger symbol denotes reversion, which can be represented by matrix transposition and change of sign of the complex unit only. The hyperbolic unit
remains unchanged. Conjugation changes in addition the sign of the hyperbolic unit.
The transition function between two charts of the frame is encoded
in terms of
\be
\label{transition}
\vartheta= e^{-\imath \Omega}\,.
\ee
The exponent on the right hand side of the equation is an abbreviation for the sixteen component tensor
\be
\label{omega}
\Omega=e_\mu\bar{e}_\nu\Omega^{\mu\nu}\,.
\ee
The tensor can be used to parameterize
the sixteen parameters of the group of general linear transformations $GL(4, \mathbb{R})$, which is acting 
on the right hand side of Eq.~(\ref{trafo}).  

Matter fields are identified with sections of the fibre bundle that is associated to the quaternion frame bundle of Sachs. 
The sections can be expanded in terms of the plane wave representation functions of the Poincar\'e group.
The transition functions are given again by Eq.~(\ref{transition}) and they are applied to the sections
in the sense of Eq.~(\ref{translation}) by multiplication from the left.

\section{Spin Spectrum of the d'Alembert Operator}
\label{wave}
 The Klein-Gordon equation encodes the free propagation of a given density distribution as
it corresponds to the inverse of the propagator or Green function, respectively. The Klein-Gordon equation
is formed by a Laplacian acting on a group manifold
minus the eigenvalue of the Laplacian with respect to the considered group representation functions.
Note that the notion of a Laplacian is applied here in a generalized sense.
Thus for the Laplacian operating on the Poincar\'e group $E_4$ and
referring  to the propagation by virtue of a four dimensional translation $T_4$ one may write
\be
\label{propagate}
\triangle(E_4\vert T_4)=p\cdot p\,.
\ee
The applied notation for the Laplacian is a generalization of the notation used by
Meng \cite{Men03} in his investigations 
of the quantum Hall effect in higher dimensions. Consider in this context also
the analysis of Hu \cite{Hu08} with respect to Laplacians in homogeneous spaces.

The notation for the Laplacian has been extended to distinguish from 
a potentially possible propagation related to the second Casimir operator of the 
Poincar\'e group, the Pauli-Lubanski vector
\be
\label{laplaceluba}
\triangle(E_4|R_4)=w\cdot w\,.
\ee
Here the propagation is driven by the Pauli-Lubanski vector $w$, which 
is defined with the help of the orbital angular momentum operators as
\be
w^\mu=\frac{1}{2}\epsilon^{\mu\nu\sigma\rho} l_{\nu\sigma}p_\rho\,.
\ee
The propagation in terms of a rotational translation may describe the dynamics of optical vortices and neutrinos.

The Laplacian defined by Eq.~(\ref{propagate}) is equivalent to the ordinary d'Alembert operator, except for a different sign. 
With the help of Eq.~(\ref{eigen}) the Laplacian can be expressed in terms
of its eigenvalues with respect to the plane wave representation functions
of the Poincar\'e group
\be
\triangle(E_4\vert T_4)=(\upsilon+\varsigma)\cdot(\upsilon+\varsigma)\,.
\ee
The dot product needs to be evaluated with Eq.~(\ref{dotprod}), which implies that the
second element in the product must be conjugated. One has to take care
that the coordinates of the spin vector coefficient change sign under conjugation
\be
\bar{\varsigma}_\mu=-\varsigma_\mu\,.
\ee
Furthermore, the vector and spin vector coefficient of the representation functions are orthogonal to each other.
Despite the new notation and interpretation the calculation can be
performed as in \cite{Ulr13}
\be
\label{spectrum}
\triangle(E_4\vert T_4)=\upsilon^2-\varsigma^2=\upsilon^2 + 4s(s+1)\upsilon^2=(\upsilon n)^2\,.
\ee 
Here Eq.~(\ref{lubadef}) and the eigenvalue of the squared 
relativistic spin angular momentum operator have been used.
The number of polarization states is derived from the spin $s$ with the formula
$n=2s+1$.
Thus one could use in Eq.~(\ref{irred}) also the spin to characterize the representation functions 
of the Poincar\'e group.
The result of Eq.~(\ref{spectrum}) can be used to formulate the following equation for fields of arbitrary spin
\be
\label{majo}
\triangle(E_4\vert T_4) \psi(x)=(\upsilon n)^2\psi(x)\,.
\ee
As mentioned in Section~\ref{intro} this equation will be denoted from now on as Majorana-Klein-Gordon equation.

A Green or propagator function, respectively, can be introduced as the inverse of the Majorana-Klein-Gordon equation with
respect to the four dimensional delta function.
The Fourier transformation of the propagator is resulting in the following expression
\be
\label{fourwave}
G_{\upsilon n}(p)=\frac{1}{p^2-(\upsilon n)^2+\imath\epsilon}\,.
\ee
The momenta are now real numbers.
The time component of the momentum vector is identified with the energy
$E=p_0$. Thus the energy spectrum can be determined from the poles of the Green function as
\be
\label{energy}
E=\pm\sqrt{\mathrm{p}^2+m^2}\,.
\ee
The three dimensional spatial momentum is denoted as $\mathrm{p}$.
The eigenvalue of the Laplacian has been parameterized in terms of the physical rest mass
\be
\label{physmass}
m^2=(\upsilon  n)^2\,.
\ee
This identification makes it possible to compare results obtained
within the Klein-Gordon method with the corresponding results of the Dirac theory. 
As an example one may consider the calculation of the Mott scattering amplitude in the Klein-Gordon fermion theory~\cite{Ulr13}. 
From the discussion in Section~\ref{matter} it 
follows that in \cite{Ulr13} the momenta of the Mott scattering amplitude have to be replaced by vector coefficients~$\upsilon$
as they are in fact derived from the Poincar\'e group representation functions. The physical momentum can then be 
reintroduced in correspondence with Eq.~(\ref{physmass})
\be
p\mapsto \upsilon=\frac{\upsilon n}{n}=\frac{p}{n}\,.
\ee
With the number of polarizations $n=2$ applied to electrons and protons
the scattering amplitude derived in \cite{Ulr13} has to be scaled down by a factor of $2^4$.  
After this transformation the Klein-Gordon scattering amplitude 
matches exactly with the corresponding result of the Dirac theory. 

\section{General Relativity}
\label{general}
Sachs has investigated a spinor representation of general relativity based on complex quaternions \cite{Sac82}.
Starting point is the Einstein-Hilbert Lagrangian with a quaternionic curvature scalar 
\be
\label{curvescalar}
{\cal L}_{EH}=R\sqrt{-g}\,.
\ee
Sachs uses a spin affine connection
to represent general relativity as
a gauge theory. The spin affine connection is introduced into the theory by the minimal substitution
\be
P_\mu=p_\mu + A_\mu\,.
\ee
The covariant derivative is defined here in terms of the momentum operator. 
The spin affine connection is given under consideration of this adjustment by
\be
\label{spinaffine}
A_\mu=\frac{1}{4}( \Gamma^\alpha_{\phantom{\alpha}\beta\mu}e^\beta+p_\mu e^\alpha)\bar{e}_\alpha\,.
\ee
The coordinate dependence of the basis elements $e_\alpha(x)$ has been surpressed for simplicity.
It should be noted that this expression has been introduced also by
Carmeli, who uses the term spinor affine connection \cite{Car01}.
The gauge transformation of the potential is defined
by Sachs as
\be
\mathcal{G}(A_\mu)=\vartheta A_\mu\vartheta^{-1} - (p_\mu\vartheta)  \vartheta^{-1}\,.
\ee
Here the transition functions introduced in Eqs.~(\ref{transition}) and (\ref{omega}) are applied.
Carmeli shows that the gauge potential is related to the Newman-Penrose spin coefficients \cite{Car01}.

The spin affine connection can be inserted as the gauge potential into the Laplacian of Eq.~(\ref{propagate})
to introduce interactions into the Majorana-Klein-Gordon equation
\be
\label{basis6}
\triangle(E_4|T_4)=(p+A)\cdot (p+A)\,.
\ee
For an explicit calculation the interaction potential can be approximated by a function.
The Laplacian then has to be diagonalized with the help of an appropriate set of eigenfunctions
\be
\triangle(E_4|T_4)\psi_\alpha(x)=c_\alpha \psi_\alpha(x)\,.
\ee
The interpretation of the spectrum in terms of known physical observables 
is not straightforward as indicated by the discussion given at the end of the previous section.

Sachs and Carmeli introduce a field strength, which is defined with respect to the gauge potential as
\be
\label{fieldtens}
F_{\mu\nu}=p_\mu A_\nu- p_\nu A_\mu +\left[A_\mu,A_\nu\right]\,.
\ee
The commutator takes care of the Yang-Mills structure of the gauge potential.
The connection coefficients in Eq.~(\ref{spinaffine}) relate the Clifford algebra of the basis elements with the Minkowski space-time coordinates.
They can be adopted from Carmeli as
\be
 \Gamma^\alpha_{\phantom{\alpha}\beta\mu}=e^\alpha\cdot P_\mu e_\beta\,.
\ee
One could introduce the notation $\Gamma^a_{\phantom{a}b\mu}$ to distinguish the indices referring to the basis elements
from the index, which refers to the covariant derivative. 
This distinction will become relevant if the gauge potential will be introduced into the Laplacian of Eq.~(\ref{conflaplace}), 
where the basis elements of the conformal solution space refer to a different dimensionality
compared to the applied differential operator.

Sachs and Carmeli are able to bring the contributions resulting from the variation of the Einstein-Hilbert action into a
Maxwell-like gauge field representation. Further Maxwell-like gauge representations of general relativity have been derived by
Rodrigues Jr. and Capelas de Oliveira \cite{Rod07} and Mielke \cite{Mie87}, see also Carmeli in \cite{Car77}.
To get an idea of this correspondence one may insert the gauge potential of Eq.~(\ref{spinaffine}) into 
the field strength tensor of Eq.~(\ref{fieldtens}). 
In relationship with gauge theories of general relativity
one may consult also the collection of research articles edited by Blagojevic and Hehl \cite{Bla13}
and compare with the representations mentioned above.
For theories with spin and torsion one may compare especially with Kibble \cite{Kib61}, Sciama \cite{Sci64}, and Hehl et al. \cite{Heh76}.

\section{Gravitation and Electroweak Interactions}
\label{interaction}
There are a number of proposals in the context of general relativity for
unified theories of gravitational and electromagnetic forces. One may have a look at
publications of Ferraris and Kijowski \cite{Fer81,Fer82a}, Hammond \cite{Ham88}, Evans \cite{Eva03}, Poplawski \cite{Pop09,Pop10}, or
Giglio and Rodrigues Jr. \cite{Gig12}. The gauge theory of Sachs can be compared with these approaches based on the following discussion.

As already mentioned the gauge transformation of the matter fields is implemented with the transition functions introduced in
Eqs.~(\ref{transition}) and (\ref{omega})
\be
\label{phase}
\mathcal{G}(\psi)=\vartheta(x)\psi(x)= e^{-\imath \Omega(x)} \psi(x)\,.
\ee
Sachs identifies gravitational and electromagnetic contributions as a dualism
within a coordinate dependent transition function~\cite{Sac78}
\be
\Omega(x)\equiv e_\mu(x)\bar{e}_\nu(x)\Omega^{\mu\nu}(x)\,.
\ee
Two limiting cases can be considered. In the first case the tensor $\Omega^{\mu\nu}$
is kept constant and the basis vectors of the Clifford algebra are
coordinate dependent $e_\mu(x)\bar{e}_\nu(x)$, which corresponds to a local variation of metric and spin
through Eq.~(\ref{basis4}). This is the Einstein-Cartan theory.
In the second limiting case the generators of the Clifford group 
$e_\mu \bar{e}_\nu$ are constant with a local variation of the tensor $\Omega^{\mu\nu}(x)$, which gives
rise to the electromagnetic part of the unified theory. 

The interpretation of the spin contributions as electromagnetism
has been discarded by Rodrigues Jr. and Capelas de Oliveira, because of Sachs' wrong interpretation
of symmetric and anti-symmetric tensor contributions within the quaternionic representation of general relativity \cite{Rod04}.
Spinor representations of general relativity are considered to be pure gravitational also by other authors. 
Carmeli interprets his representation in this sense and 
adds additional internal symmetries to include electromagnetic or Yang-Mills fields, respectively \cite{Car01}.
However, the gauge transformation of Eq.~(\ref{phase}) has not been considered in these two references.
Blagojevic and Hehl have made a general statement  
on fallacies about torsion with respect to unified theories of
gravitation and electromagnetism \cite{Bla13}. Sachs does not identify electromagnetism with torsion, but with the matrix structures inside metric
and spin tensor, see again Section \ref{para} for more details on these matrix structures.

Electromagnetism in its common understanding results from a one parameter gauge group. 
It should be noted however that Sachs considers in fact an electroweak theory and uses the notion of
electromagnetism in a unified sense. The method is applied for example to the weak decay of neutrons \cite{Sac86}.
The electroweak interactions are described within the Glashow-Salam-Weinberg theory as a four parameter gauge group.
The corresponding gauge symmetry appears also within the algebra of Sachs as the spatial basis elements $e_i$ satisfy
the $SU(2,\mathbb{C})$ Lie algebra and $e_0=1$ gives
rise to a $U(1,\mathbb{C})$ gauge symmetry.
In this context it is remarkable that in a simple Yukawa parameterization the coupling constants of
electromagnetic and weak interactions have the same magnitude
and only the range of the interactions is different due to the different masses of the exchanged vector bosons.
Consider Nachtmann for more details~\cite{Nac86}.

The product of basis elements $e_\mu \bar{e}_\nu$, which appears in the definition of the gauge transformation, forms the Lie algebra of the Pin group,
see G\"ockeler and Sch\"ucker \cite{Goc87}. In this case it is the group $Pin(1,3)$.
Thus the electroweak interaction can be understood in this representation as being related to the following group structure
\be
\label{electroweak}
G_{EW}=Pin(1,3)\cong U(2)^2=(SU(2)\times U(1))^2\,.
\ee
The unitary groups are defined over the field of complex numbers~$\mathbb{C}$.
The squared group has to be understood as
\be
\label{squaregroup}
G^2=G\times G\,.
\ee
Alternatively, one can drop the square and consider the group as being defined over the ring of hyperbolic complex numbers $U(2,\mathbb{H})$
or identify it as the general linear group $GL(2,\mathbb{C})$. 
The notation $\mathbb{H}$ stands for the combination of complex
and hyperbolic numbers, which can be found in the literature also as bicomplex numbers of Segre or tessarines of Cockle.

Note that there is an alternative identification of gravitational and non-gravita\-tional contributions in the gauge group.
Gravitation could appear already in flat space-time as a hyperbolic complex Maxwell theory~\cite{Ulr06}.
Thus gravitation would be a substructure in the $U(2,\mathbb{H})$ gauge symmetry mentioned above. 
In this picture the gauge symmetry would include the electroweak interaction 
as a further substructure beside the gravitational interaction on the same mathematical level.
However, the suggested interpretation would imply a gravitational counterpart of the weak interaction to complete the $U(2,\mathbb{H})$ gauge symmetry.
This gravitoweak interaction has not been observed in experiment so far. Therefore the notation $G_{EW}$ in Eq.~(\ref{electroweak}) will
not be adjusted with respect to this interpretation.

\section{Conformal Compactification of Minkowski Space}
\label{confalg}
The methodology discussed in the previous sections is situated in the non-compact Minkowski space.
One possibility to enclose the unlimited space-time geometry is to add infinity by virtue of
the conformal compactification, see Penrose and Rindler \cite{Pen86}.
In the context of Clifford algebras this procedure is described for example by Porteous \cite{Por95,Por96}.
The Minkowski space is extended to six dimensions, where the first four dimensions are identical $u^\mu=x^\mu$.
Furthermore, two additional coordinates are introduced
\be
u^4=\frac{x\bar{x}+1}{2}\,,\quad u^5=\frac{x\bar{x}-1}{2}\,.
\ee
The square of the conformal vector
will result in a null vector if the metric $\mathbb{R}^{2,4}$ is applied and
$u^4$ is assigned to the additional negative sign coordinate with respect to this metric.

Conformal structures and twistors in the paravector model of space-time have been investigated before by Elstrodt, Grunewald, and Mennicke \cite{Els87},
Maks \cite{Mak89}, and da Rocha and Vaz \cite{Roc07}.
It is thus natural to reintroduce this formalism under consideration of the hyperbolic complex number system. 
The basis elements for a paravector in conformal space $e_\mu=(1,e_i)$ are then defined in terms of the Dirac algebra, which 
corresponds to the Clifford algebra $\mathbb{R}_{4,1}$. 
In analogy to the hyperbolic Pauli algebra, 
the notion of a hyperbolic Dirac algebra is introduced to label the hyperbolic complex representation of the Dirac algebra.
Note again that the hyperbolic Dirac algebra is isomorphic to the ordinary Dirac algebra.
The first three non-trivial basis elements are
\be
\label{basspace}
e_{i}=
\left(\begin{array}{cc}
j\sigma_i&0\\
0&-j\sigma_i
\end{array}\right)\,.
\ee
The matrix on the right hand side of the equation is formed by the basis elements of the hyperbolic Pauli algebra, see Eq.~(\ref{Pauli}).
Two additional matrices have to be introduced to match the $\mathbb{R}^{2,4}$ metric of conformal space
\be
\label{basrest}
e_4=\left(\begin{array}{cc}
0&ij\\
-ij&0
\end{array}\right)\,,\quad
e_5=\left(\begin{array}{cc}
0&i\\
i&0
\end{array}\right)\,.
\ee
Based on these definitions the conformal vector can be represented as a paravector in formal analogy with the Minkowski vector
$u=u^\mu e_\mu$.

Most of the formulas in the first sections remain valid due to this formal analogy.
They can be applied also in conformal space under consideration of the corresponding dimension, metric, and basis elements.
The antisymmetric contributions $\sigma_{\mu\nu}$ are defined again by Eq.~(\ref{spinprod}),
but now the spin tensor has a $6\times 6$ matrix structure
and each of the matrix elements is a $4\times 4$ matrix in its fundamental representation.
The generators of the spin angular momentum are given again by Eq.~(\ref{spin}),
but now calculated with the basis elements of the conformal algebra
\be
\label{polten}
s_{\mu\nu}=\frac{\sigma_{\mu\nu}}{2}\,.
\ee
In analogy to Kastrup \cite{Kas62} one can define with these generators
spin representations of momentum, angular momentum,
special conformal transformations, and scale transformations.
The spin tensor is furthermore eligible for higher spin representations.

The pseudoscalar of the conformal algebra $\mathbb{R}_{4,1}$, which is understood as the oriented volume element,  
is calculated under consideration of the six basis elements.
This leads to a different result compared to the corresponding expression in Minkowski space
\be
\label{pseudoconf}
\imath=e_0\bar{e}_1e_2\bar{e}_3e_4\bar{e}_5=i\;.
\ee
The hyperbolic unit is not included in the volume element of conformal space because 
four dimensions in the metric appear with a negative sign. To each of these four dimensions a hyperbolic unit is assigned
and they finally cancel each other by multiplication. 
In order to distinguish the different volume elements,
 the pseudoscalar of Minkowski space defined in Eq.~(\ref{pseudo}) is abbreviated in the remainder of this section by~$\imath_E$ 
and the pseudoscalar of conformal space
by~$\imath_P$
\be
\imath_E=ij\,,\quad\imath_P= i\,.
\ee
The square of the volume elements is obtained, if
they are multiplied in each of the algebras by their conjugated element
\be
\imath_E\bar{\imath}_E=-1\,,\quad\imath_P\bar{\imath}_P=+1\,.
\ee
One may say that the volume element is negatively charged
in the first case and positively charged in the second case. Further multiplication of these products leads to
\be
\imath_E\bar{\imath}_E\imath_P\bar{\imath}_P=-1\,.
\ee
Here the notion of attraction with respect to the volume elements can be introduced. The corresponding product with only $\imath_E$
or $\imath_P$ pseudoscalars is leading to a positive sign and thus repulsion.

\section{Conformal Plane Waves}
\label{confsol1}
Plane wave representation functions of the conformal group 
can be introduced in analogy to the representation functions of
the Poincar\'e group. The conformal group space is strictly spoken extended by translations to provide
a formal analogy to the  Poincar\'e group wave functions. Nevertheless, the notion of conformal group will be used for simplicity in this context. 
The wave functions now have the form
\be
\label{confwave}
\psi(u)=\exp{(-\imath u \bar{\gamma})}\,.
\ee
The conformal compactification of a Minkowski vector $u=u^\mu e_\mu$, expressed 
with the coordinates and basis elements introduced in the previous section,
can also be written as
\be
\label{confcoord}
u=\left(\begin{array}{cc}
x&\alpha(x)\\
-\bar{\alpha}(x)&\bar{x}
\end{array}\right)\,.
\ee
The conformal compactification $u\equiv u(x)$ is a matrix valued function of the Minkowski Clifford paravector $x$.
The non-diagonal elements are abbreviated in terms of the function
\be
\label{Kappa}
\alpha(x)=x\bar{x}o+\bar{o}\,.
\ee
Beside the squared Minkowski vector the function includes a complexified
representation of the null plane numbers, which have been applied before for example by Zhong \cite{Zho85} and Hucks \cite{Huc93}
\be
o=\frac{i+ij}{2}\,.
\ee
The calculation rules for the complexified null plane numbers can be compared with the 
ordinary null plane numbers
\be
\label{nullcalc}
oo=\imath o\,,\quad \bar{o}\bar{o}=-\imath\bar{o}\,,\quad o\bar{o}=0\,.
\ee
They differ in the pseudoscalar on the right hand side of the first two equations.
The compactification of the vector coefficient $\upsilon$, which labels the Poincar\'e group representation functions, can be represented
in analogy to Eq.~(\ref{confcoord})
\be
\label{confvec}
\gamma=\left(\begin{array}{cc}
\upsilon&\alpha(\upsilon)\\
-\bar{\alpha}(\upsilon)&\bar{\upsilon}
\end{array}\right)\,.
\ee
Together with the pseudoscalar 
of conformal space the conformal plane waves of Eq.~(\ref{confwave}) are then defined in detail.

The conformal vector given by Eq.~(\ref{confcoord}) has been represented in the context of 
M\"obius transformations in a similar form by Vahlen \cite{Vah02}, Ahlfors \cite{Ahl86}, 
Fillmore and Springer \cite{Fil90}, and Cnops \cite{Cno94,Cno02}.
In the terminology of Kisil \cite{Kis12b}, which is thought to be generalized accordingly to Minkowski space, the conformal vector
is classified as a $h$-zero radius cycle within the geometries of the Erlangen program
of Klein. The $h$ stands for a hyperbola. The notion of a cycle combines the geometric objects of 
circles, parabolas, and hyperbolas as conformal invariant conic sections. These three cases are distinguished
by their corresponding hypercomplex units, which square to $-1$, $0$, and $1$. 
The terminology of a cycle traces here back to Yaglom \cite{Yag79}. It has been extended and adjusted by Kisil \cite{Kis12b}. 
The zero radius hyperbola is interpreted 
geometrically as a light cone with center at $x$, see for example Kisil in \cite{Kis07}. Thus the
mathematical structures resulting from the conformal compactification are elements of a cycle space. The cycles can be
transformed by M\"obius transformations and they are displayed in a point space, which corresponds to
the four dimensional Minkowski space.

\section{The Wave Equation}
\label{confsol2}
It is now possible to calculate the spectrum of the Laplacian with respect to the wave functions
introduced in the previous section
\be
\label{conflaplace}
\triangle(C_4\vert T_4)=p\cdot p\,.
\ee
Here the short notation $C_4$ has been introduced to indicate that the Laplacian is operating on the
representation functions of the conformal group. The argument in the exponential of the group representation functions
is determined by cycles, as discussed in the previous section. The cycles are displayed in the four dimensional 
Minkowski point space. Consequently, the momentum operator remains a four dimensional 
vector derivative acting in the Minkowski point space.
The Green function, which is derived from the Laplacian, refers to the propagation of a density distribution by virtue of a four dimensional translation. 
This is indicated by the notation $T_4$. 
Compared to Eq.~(\ref{propagate}) the momentum operator is
defined with the volume element of the conformal space as introduced in Eq.~(\ref{pseudoconf}). 

The Laplacian is acting on the conformal plane waves of Eq.~(\ref{confwave}).
As an alternative to the calculation which leads to Eq.~(\ref{spectrum}) one can operate directly
with the four dimensional Minkowski space derivative on the conformal plane waves, in detail
\be
p_\mu \psi(u)= \left(\frac{\partial}{\partial x^\mu}u\bar{\gamma}\right) \psi(u)\,.
\ee
To calculate the spectrum of the Laplacian the momentum operator has to be applied twice according to Eq.~(\ref{conflaplace}).
One has to take care that the matrix valued vector obtained from the second derivative in the Laplacian
has to be conjugated, which follows from the definition of the scalar product in Eq.~(\ref{dotprod}).
The resulting contributions cancel each other 
and one finds
\be
\label{confwaveeq}
\triangle(C_4\vert T_4)\psi(u)=0\,.
\ee
At first sight this result seems to be not very surprising as the compactified vector coefficient is a null vector in conformal space
and the conformal plane wave is thus expected to refer to  light like conformal group representation functions.
However, the operator acting on the conformal plane wave corresponds up to a different sign to the ordinary d'Alembert operator.
In other words, the dimensionality differs between the
four dimensional Minkowski space of the differential equation, which can be labelled as propagation space, and the solution space.
In this case the solution space is  the six dimensional conformal space, which reflects the symmetries of the 
Laplacian as the underlying differential equation.

The conformal plane waves thus satisfy the wave equation for all higher spin representations and masses,
as they appear within the Poincar\'e subgroup. 
To see the above solutions in a wider context one may compare here with
the conformal zero rest mass fields of arbitrary spin, which have been investigated by Penrose \cite{Pen65,Pen68}
and Penrose and MacCallum \cite{Pen73}.
With respect to the differential operator it should be noted that
the idea to trace quantum physics back to the wave equation has been suggested before by Sallhofer \cite{Sim05}
in the context of the Schr\"odinger theory. 

In Section \ref{axiom} it will be argued that the extension to conformal symmetries induce the strong interaction.
The electroweak interactions have been positioned in Section~\ref{interaction}
within a reduced solution space, where the corresponding differential equation is only invariant with respect to the Poincar\'e transformations. 
Spontaneous and anomalous symmetry breaking may
reduce the symmetry of the overall mathematical structure and makes mass and spin visible.
Equation~(\ref{confwaveeq}) is free of these parameters and thus provides a more general view.
One may therefore select the wave equation for the development of a relativistic field theory.
The wave equation, as the inverse of the propagator, 
can be understood as the mathematical representation of Newton's first law of motion applied to fields.
The propagation rule of Newton is the conceptual source of the method.

The next step is to impose the principle of gauge invariance.
This results in the introduction of a gauge potential, which modifies the wave equation according to Eq.~(\ref{basis6})
\be
\label{basis7}
\triangle(C_4|T_4)=(p+A)\cdot (p+A)\,.
\ee
Here the spin affine connection of Eq.~(\ref{spinaffine}) has been applied to provide a geometric field theory. Note that the basis
elements $e_\mu$ of the conformal solution space can be used in the definitions of Section~\ref{general}.
The gauge potential implements the invariance of the solutions with respect to gauge transformations,
which are induced by the global symmetries of the underlying differential equation.
Thus the principal bundle is naturally defined. 
It couples the differential equation, the propagation rule for densities defined in the
base manifold of the bundle, with the bundle structure group that refers to the symmetries of the differential equation.

\section{Strong Interactions and Skyrmions}
\label{axiom}
Since the sixties of the last century various models have been proposed, which consider
conformal symmetries to understand the structure of the strong interaction, see the discussion in Section~\ref{intro}.
In this sense one may extend the gauge theory of Sachs to conformal space
and search for indications which can be interpreted as an extension of the theory of 
electroweak interactions to a theory which includes also the strong interaction.
The gauge transformation in the conformal solution space is formally defined as in Eq.~(\ref{phase})
\be
\mathcal{G}(\psi)=\vartheta(u)\psi(u)= e^{-\imath \Omega(u)} \psi(u)\,.
\ee 
The exponent $\Omega$ is given formally by Eq.~(\ref{omega}), but now
the basis elements of the conformal Clifford algebra $\mathbb{R}_{4,1}$
introduced in Eqs.~(\ref{basspace}) and (\ref{basrest}) are applied.
The tensor elements $e_\mu \bar{e}_\nu$ form the Lie algebra of the Pin group,
which corresponds now to
\be
\label{pati}
G_{PS}=Pin(2,4)\cong U(4)^2 \,.
\ee
The unitary group is defined over the field of complex numbers~$\mathbb{C}$.
The square of a group is defined as in Eq.~(\ref{squaregroup}). Again one may drop the square and
consider the group instead as being defined over the ring of hyperbolic complex numbers~$\mathbb{H}$
or identify it as the general linear group $GL(4,\mathbb{C})$.
The group $U(4)\times U(4)$ corresponds to an extension of the Pati-Salam model \cite{Pat73}.
The Pati-Salam model is a so-called preon model, which postulates the existence of substructures of quarks.
The $U(4)\times U(4)$ symmetry appears explicitly in the unified gauge theory of conformal gravity, Maxwell
and Yang-Mills fields of Castro \cite{Cas11,Cas12}. It forms furthermore a slight extension of the model of Fayyazuddin~\cite{Fay12}. 
The $U(4)$ symmetry is considered in the gauge invariant spinor theory of D\"urr~\cite{Due81}.

Alternatively, one can assign linear combinations of the thirty-two elements of
the Clifford algebra $\mathbb{R}_{4,1}$ to the
group generators of
\be
\label{han}
G_{GM}=(SU(3)\times SU(3))^2\,.
\ee
This partition reminds of quark models proposed by Gell-Mann \cite{Gel62,Gel68} and Han and Nambu \cite{Han65},
furthermore to the chiral symmetry of Dashen \cite{Das69}. 
One may continue to break down the conformal algebra into a thirty-two dimensional 
Lie algebra, which resembles the Standard Model
\be
\label{standard}
G_{SM}=(SU(3)\times(SU(2)\times U(1))^2)^2\,.
\ee
Note that Wheeler was able to identify the Standard Model gauge group as a residual symmetry
of extended conformal gravity \cite{Whe92}. 

Matrix algebras of multiple $SU(2)$ and $U(1)$ groups can be related to the thirty-two dimensional conformal Clifford algebra
\be
\label{ser}
G_{QHD}=(SU(2)\times U(1))^4=U(2)^4\,.
\ee
The partition into these smaller gauge groups may correspond to models in quantum hadrodynamics (QHD).
Consider Serot and Walecka \cite{Ser86} for more details on the representation of the strong interaction in the low energy regime
in terms of superstructures of quarks. 

One may think also of an extension of the $SU(2)\times SU(2)$ chiral Skyrme model \cite{Sky61},
whose solitons can be interpreted as the baryons of QCD, see Witten \cite{Wit83a,Wit83b} and Adkins, Nappi, and Witten \cite{Adk83}.
For example Pomarol and Wulzer extended to a $U(2)\times U(2)$ gauge symmetry to describe baryons
as Skyrme-like solitons \cite{Pom09}.
Ma et al. investigated an $U(2)$ extension of the original Skyrme model, where the extension is understood as a hidden local symmetry \cite{Ma13}.
More background on Skyrmions and solitons can be found in 
Alkofer and Reinhardt \cite{Alk95}, Manton and Sutcliffe \cite{Man04}, Brown and Rho \cite{Bro10}, Weigel \cite{Wei08}, and Dunajski \cite{Dun10}.
The investigations which focus on effective meson models are founded on the large number of colors limit of 't Hooft \cite{Hoo73}. 
One may ask now whether the description can be extended beyond pure strong interactions in accordance with the above gauge groups.
This question is of importance if also leptons are interpreted as solitons, see for example Weiner \cite{Wei13}.
In this context one may consider the geometric models of matter of Atiyah, Manton, and Schroers \cite{Ati12},
who describe electrons, protons, neutrinos, and neutrons with a method that has been inspired by Skyrme's baryon theory.

The correspondence between the gauge groups mentioned in this section has to be understood in the following sense. 
The generators of the subgroups can be represented in terms of generators of
the higher dimensional groups. For example there are three possibilities to express the generators of $U(2)=SU(2)\times U(1)$
in terms of the $SU(3)$ Gell-Mann matrices. One of them is up to factors given by
\be
\mathfrak{g}(U(2))=\{\lambda_1,\lambda_2,\lambda_3,\lambda_8\}\,.
\ee
Similarly, one can represent the generators of the considered subgroups with linear combinations of $U(4)$ generators.
With the help of these representations the full group space can be spanned in accordance with Eq.~(\ref{pati})
starting from Eqs.~(\ref{han}), (\ref{standard}), or (\ref{ser}).

The discussion indicates that the gauge field structures resulting from the conformal compactification 
have the potential to describe the strong interaction.
The partition of the Pin group into subgroups is not unique. Accordingly, there are different theoretical
models which explain manifestations of the strong interaction depending on the considered energy region.

\section{Summary}
The Majorana-Klein-Gordon equation is a hyperbolic complex second order differential equation
describing the dynamics of matter fields with arbitrary spin. The equation is embedded into the gauge field theories of 
Sachs and Carmeli, where Sachs considers a unified representation of general relativity and electroweak interactions. 

The method can be generalized to conformal space with the intention to incorporate the strong interaction. 
The conformal compactification of a Minkowski vector can be interpreted as a light cone,
which is still situated in Minkowski space. It is then possible to trace relativistic physics back to the wave equation
acting on representation functions of the conformal group.

\end{document}